\documentclass[aps,prl,preprint]{revtex4}
\usepackage{mathrsfs}
\usepackage{graphicx}
\usepackage{amssymb}
\usepackage{setspace}

\begin{document}

\title{Resolution and enhancement in nanoantenna-based \\fluorescence microscopy}

\author{Hadi Eghlidi, Kwang Geol Lee, Xue-Wen Chen, Stephan
G\"{o}tzinger, and Vahid Sandoghdar }


\affiliation{Laboratory of Physical Chemistry and optETH, ETH
Zurich, CH-8093 Zurich, Switzerland}


\begin{abstract}
Single gold nanoparticles can act as nanoantennas for enhancing the
fluorescence of emitters in their near fields. Here we present
experimental and theoretical studies of scanning antenna-based
fluorescence microscopy as a function of the diameter of the gold
nanoparticle. We examine the interplay between fluorescence
enhancement and spatial resolution and discuss the requirements for
deciphering single molecules in a dense sample. Resolutions better
than 20~nm and fluorescence enhancement up to 30 times are
demonstrated experimentally. By accounting for the tip shaft and the
sample interface in finite-difference time-domain calculations, we
explain why the measured fluorescence enhancements are higher in the
presence of an interface than the values predicted for a homogeneous
environment.

\end{abstract}

\maketitle

Since its realization in $1984$ \cite{Pohl1984,Lewis1984}, Scanning
Near-field Optical Microscopy (SNOM) has demonstrated spatial
resolution beyond the diffraction limit in several different
contexts and configurations. Some of the most impressive reports of
resolution have been made in the so-called apertureless SNOM mode
where the intensity of a diffraction-limited illumination is
enhanced in a small nanoscopic region very close to a sharp metallic
tip~\cite{Zenhausern:94,gleyzes:95,kawata:95,zenhausern:95,Sanchez:99,Hartschuh:08}.
Despite its success in nonlinear~\cite{Hartschuh:08} and
interferometric infrared~\cite{Keilmann:04} near-field imaging,
apertureless fluorescence SNOM has not become a routine tool in the
laboratory. Moreover, theoretical modeling of the optical properties
of tips made of real metals has posed numerous challenges so that
quantitative agreements between theory and experiment have been
missing. Several years ago, we proposed to use single spherical gold
nanoparticles (GNP) attached to dielectric tips as well-defined and
reproducible probes for apertureless SNOM~\cite{Kalkbrenner:01}.
Since then it has been shown that a GNP can act as a nanoantenna to
enhance the fluorescence of a single molecule (SM) in its near
field~\cite{Anger:06,Kuehn:06,Kuehn:08,Hoeppener:08}. The simplicity
of this experimental arrangement has made it possible to compare the
experimental findings with theoretical predictions.

As the GNP becomes smaller, the region to which its near-field
intensity is confined shrinks, suggesting the possibility of
achieving higher spatial resolutions. However, to benefit from this
strong confinement, the molecule-GNP separation also has to be made
smaller and this might in turn cause fluorescence
quenching~\cite{Kuehn:06b,Anger:06,Kuehn:08}. Viewed from a
different perspective, smaller GNP antennas are expected to become
less effective because the ratio of the scattering and absorption
cross sections decreases with decreasing GNP
diameter~\cite{Kreibig}. In this Letter, we examine the intricate
interplay between excitation enhancement, quenching, and resolution
as a function of the GNP diameter.

\begin{figure}[h]
\includegraphics[width=0.4\textwidth]{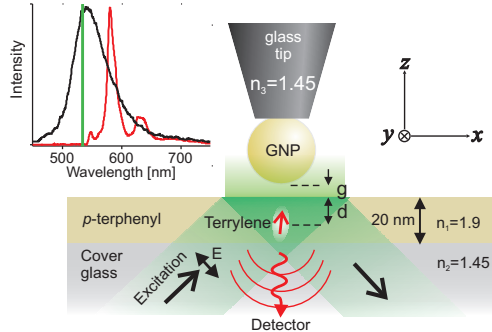}
\caption{Sketch of the experimental setup. Terrylene molecules are
embedded in a thin crystalline \emph{p}-terphenyl film and are
illuminated under total internal reflection. The same objective
collects fluorescence of individual molecules while a single gold
nanosphere attached to a glass fiber tip is scanned across.
Parameters $d$ and $g$ denote the depth of the molecule under the
sample surface and the GNP-sample separation, respectively.
Refractive indices n$_{1}$, n$_{2}$, and n$_{3}$ show the values
used for the theoretical calculations. The inset shows a plasmon
spectrum of a GNP (black), the fluorescence spectrum of a single
terrylene molecule (red), and the excitation laser line (green).}
\label{setup}
\end{figure}

Our samples consisted of terrylene molecules embedded in ultrathin
crystalline films of \emph{p}-terphenyl (pT) spin coated on
carefully-cleaned glass cover
slides~\cite{Pfab:04,Kuehn:06,Kuehn:08,Kuehn:06b}. In these films,
terrylene molecules are aligned with their transition dipole moments
at about $15^\circ$ to the normal to the plane of the sample. We
excited the molecules by light from a solid-state laser at a
wavelength of $\lambda=532$~nm through a high numerical aperture
microscope objective (N.A.$=1.4$). As shown in Figure~\ref{setup},
the excitation laser beam was p-polarized to generate a strong
$z$-component upon total internal reflection at the pT-air
interface. Fluorescence from individual molecules, which was peaked
at $\lambda=580$~nm, was collected with the same objective and
directed either to a sensitive charge-coupled device (CCD) or to a
single photon counting avalanche photodiode (APD). The fluorescence
signal of a single molecule was then monitored while a GNP was
laterally raster scanned across it at a constant gap $g$ from the
pT-air interface. The details of tip preparation, tip-sample
distance stabilization, as well as excitation and detection schemes
can be found in
Refs.~\cite{Kalkbrenner:01,Kuehn:06,Kuehn:08,Kuehn:06b}.

\begin{figure}[h!]
\includegraphics[width=0.46\textwidth]{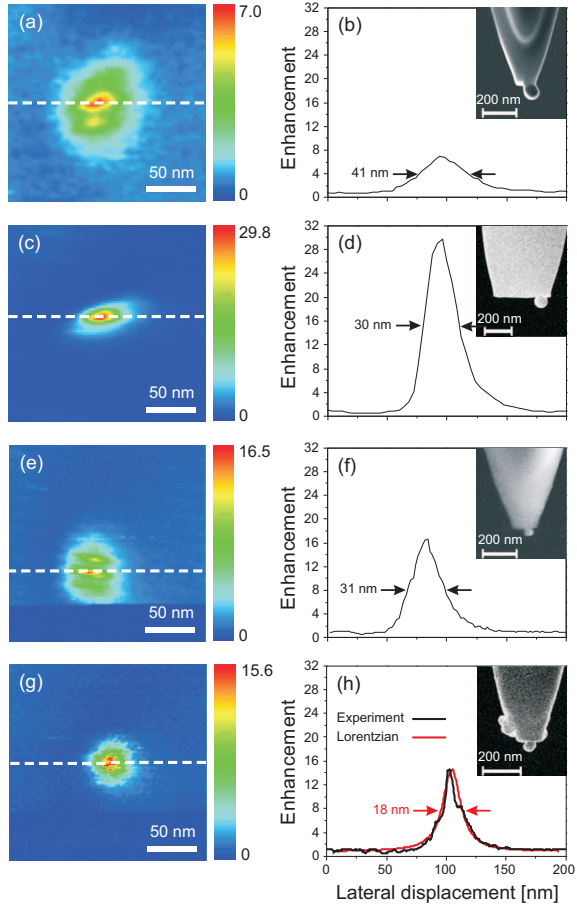}
\caption{Near-field fluorescence images of single molecules obtained
using gold nanospheres with diameters of 100~nm (a), 80~nm (c),
60~nm (e), and 40~nm (g). The enhancement is encoded in false
colors. The images were recorded at a scan speed of 40~nm/s with a
pixel size of 2~nm. The squeezed shape of the image in (c) is caused
by a stage drift. b, d, f, h) Cross sections of the scan images on
the left-hand side. The red curve in (h) displays a Lorentzian fit.
The insets show SEM images of the gold nanospheres attached to the
fiber tips.} \label{sizes}
\end{figure}

Figure~\ref{sizes}a shows a $200$\,nm x $200$\,nm raster scan
fluorescence image of a single terrylene molecule under a GNP with a
diameter of 100~nm. In Figure~\ref{sizes}b, we plot a cross section
from this image, reporting a full width at half-maximum $W=41$~nm
and a fluorescence enhancement $\Phi=8$. An electron micrograph
(SEM) of the probe is displayed in the inset. To determine $\Phi$
for each SM, we carefully measured the sample fluorescence
background from a region without the molecule and subtracted it from
the fluorescence signals of the molecule in the presence and absence
of the GNP. Furthermore, we accounted for the autofluorescence of
each GNP. Figures~\ref{sizes}c-h present some examples of the
results from experiments with GNPs that had diameters 80, 60, and
40~nm. Figures~\ref{sizes}c,d report $\Phi$ as large as $30$ while
Figures~\ref{sizes}g,h show that $W$ can be as narrow as $18$\,nm.
Interestingly, $\Phi=16$ for a GNP with a diameter of 40~nm, which
is even larger than the value obtained in Figures~\ref{sizes}a,b for
a GNP that is 100~nm in diameter.

\begin{figure}[]
\includegraphics[width=0.4\textwidth]{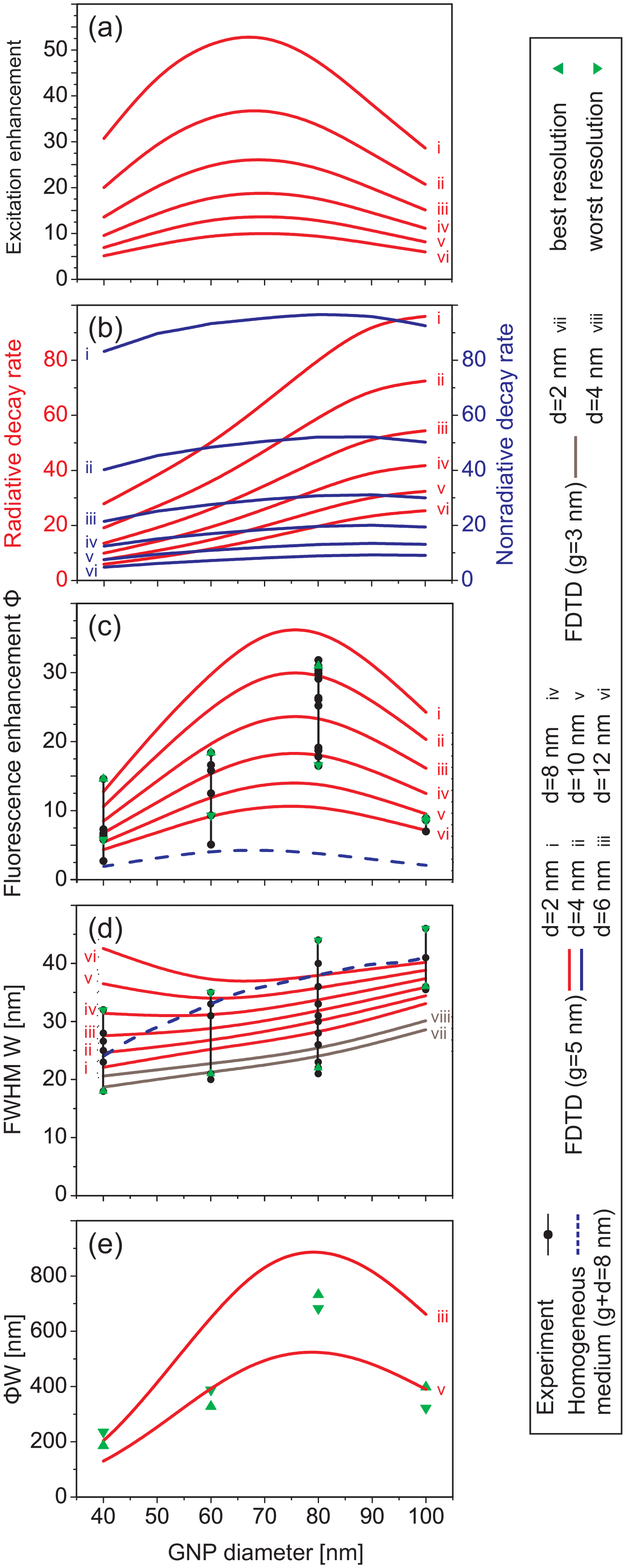}
\caption{ a) Enhancement of the excitation intensity, b) radiative
(red curves) and nonradiative (blue curves) decay rates normalized
to the radiative decay rate in the absence of the GNP, c)
fluorescence enhancement $\Phi$, d) full width at half-maximum $W$,
e) the product $W\Phi$. (a)-(e) are studied as a function of the GNP
diameter. The black circles display the experimental measurements on
single molecules. The solid curves show the results of the FDTD
calculations for different molecule depths $d$ and tip-sample
separations $g$ as specified in the legends; In (a)-(c) we employed
bor-FDTD whereas in (d) we performed full 3D calculations to find
the quantum efficiencies of laterally displaced molecules. The blue
dashed curves depict the results of analytical calculations in a
homogeneous medium of refractive index 1.35. The green triangular
symbols pointing up and down denote the data corresponding to the
highest and lowest observed resolutions, i.e. lowest and highest
values of $W$, respectively.} \label{ex-theory}
\end{figure}

To identify a size-dependent trend, we performed experiments with
many antenna probes and SMs. Instead of reporting average values, we
present the outcome of individual measurements for $\Phi$ and $W$ by
the black symbols in Figures~\ref{ex-theory}c and d, respectively.
For convenience, we have labeled the lowest and highest values of
$W$ for the data set of each particle size by the upward and
downward pointing triangles. As expected, in most cases a low $W$ is
correlated with a high $\Phi$ and vice versa. To examine the origin
of the substantial spread in our findings, we now present the
results of finite-difference time-domain (FDTD) calculations with
convolutional perfectly matched layer absorbing boundary
condition~\cite{Taflove}. To minimize the computation time, we used
the body of revolution FDTD method (bor-FDTD) whenever
applicable~\cite{Mohammadi:08}. In particular, we made the
approximation that the molecular dipole moments are perpendicular to
the pT film to benefit from the bor-FDTD calculations. The great
advantage of this numerical approach over a simple analytical
model~\cite{Das:02, Kuehn:06} is that it allows us to take into
account the tip shaft, pT thin film, underlying glass substrate, as
well as the evanescent excitation.

In the weak excitation regime, $\Phi$ is the product of the
enhancement in the excitation intensity, modification of the quantum
efficiency (QE), and change in the collection efficiency due to the
presence of the GNP~\cite{Kuehn:06}. To find the first factor, we
propagated a radially polarized Gaussian beam incident at an angle
of $45$ degrees in the bor-FDTD calculations to mimic a p-polarized
illumination used in the experiment (see Figure~\ref{setup}). We
then calculated the electric field intensity at different locations
inside the pT film in the presence and absence of the antenna probe.
Figure~\ref{ex-theory}a displays some examples of the outcome for
$g=5$~nm and different values of $d$ as a function of the GNP
diameter. To evaluate the QE of the emitter for a given
configuration, we computed the power radiated by the molecule into
the far field as well as the total power dissipated (i.e.
radiatively and nonradiatively) by the molecule and determined the
ratio of the former to the latter. The red and blue curves in
Figure~\ref{ex-theory}b show the calculated radiative and
nonradiative decay rates normalized to the spontaneous emission rate
of the molecule in the absence of the nanoantenna. The collection
efficiency was calculated as the ratio of the radiated power into
the substrate within a cone angle of $135$ degrees to the total
radiated power.

The solid curves in Figure~\ref{ex-theory}c summarize the results of
computations for $\Phi$, indicating that the spread in $\Phi$ can
stem from small variations of the distance $d$ between the molecule
and the pT-air interface (see Figure~\ref{setup}). However, since
the tip-sample distance in SNOM is not stabilized to an absolute
value~\cite{Karrai-00PRB}, it is also possible that the fluctuations
are caused by small changes in $g$. An extended study of the various
parameters such as $d$, $g$, and the refractive indices of the film
and the substrate go beyond the scope of this paper and will be
discussed in a forthcoming publication~\cite{Chen:09}.

A sensitive dependence of $\Phi$ on the molecule-GNP separation is
also expected from an analytical model of a molecule close to a GNP
embedded in a homogeneous medium~\cite{Das:02,Kuehn:06}. However,
the outcome of such a simple analysis, shown by the blue dashed
curve in Figure~\ref{ex-theory}c for a molecule-GNP separation of
8~nm (equivalent to $g+d$), does not explain the observed
fluorescence enhancements as large as 30 times~\cite{note2}. As we
had anticipated in our previous work~\cite{Kuehn:06}, a quantitative
understanding of $\Phi$ can only be obtained if one takes into
account the sample-air interface. One of the effects of the
interface is to lengthen the radiative lifetime of the excited state
for a molecule with a dipole oriented normal to the
interface~\cite{Buchler:05}. For terrylene, which embeds in pT films
at an angle of about $15^\circ$, this amounts to 4-5 fold reduction
of spontaneous emission rate compared to its bulk value. Thus, the
quantum efficiency of these molecules is reduced from about 95\% to
79\% \cite{note}. Calculations reveal that the onset of quenching is
delayed by the interface so that for an axially-oriented molecule
placed at $d=2$~nm, $g=2$~nm, the quantum efficiency is as high as
30\% as compared to 15\% for a GNP-molecule separation of 4~nm in a
homogeneous environment with a refractive index of
1.35~\cite{Chen:09}. Furthermore, the presence of the interface
assists the antenna-induced enhancement of the excitation intensity
and therefore $\Phi$.

The data in Figure~\ref{ex-theory}c show that fluorescence
enhancement reaches an optimum for GNP diameters around 80~nm while
$\Phi$ drops to values of the order of unity for GNP diameters of
40~nm and below. The solid curves in Figure~\ref{ex-theory}d display
the expected trend of an improved resolution for smaller GNPs and
shorter molecule-GNP separations. These data were obtained by
calculating $\Phi$ for different lateral positions of the molecule
and then assessing the FWHM of the fluorescence profile. Here, one
has to keep in mind that in order to achieve higher resolutions with
smaller GNPs, it is also imperative to reduce the molecule-GNP
separation. The rise of $W$ observed in curves iv, v, and vi for
small GNPs illustrates this point. The blue dashed curve shows that
as opposed to the case of $\Phi$, the magnitude and trend of $W$ do
not deviate by a great deal from the predictions of an analytical
model for a homogeneous medium with a refractive index of 1.35.

With our current samples, it was not possible to ensure GNP-molecule
distances of only a few nanometers in a routine fashion. A
theoretical scrutiny of this regime is also beyond our current work
because of numerical difficulties as well as uncertainties in the
knowledge of the dielectric functions of very small gold
nanoparticles~\cite{Kreibig,Stoller:06}. Furthermore, we point out
that in cases where quenching becomes important, it is a nontrivial
task to determine $W$ because lateral cross sections could first
show enhancement and then quenching, leading to complex line
shapes~\cite{Anger:06,Kuehn:08}. Indeed, in the $\Phi<1$ regime,
quenching has to be used as a contrast mechanism for
microscopy~\cite{Yoskovitz:08}. Finally, we point out that
antenna-based microscopy is most efficient for imaging axial dipole
moments as studied in this work, and that the competition between
enhancement and quenching is very different for emitters with a
transition dipole moment parallel to the substrate~\cite{Kuehn:06b}.

The ultimate task of any imaging technique is to describe unknown
samples. It is, thus, important to examine the ability of
antenna-based microscopy for resolving close-lying fluorophores in a
dense surrounding. Let us consider the fluorescence signal of a
single emitter to be $S_0$ counts within the integration time. A
uniform distribution of emitters at surface density $\delta$
molecules/$\rm nm^2$ in a diffraction-limited detection spot yields
a background signal of $S_0 \delta\lambda^2 /(2N.A.)^2 $, whereas
the signal from the enhanced region reads $S_0\Phi\delta W^2$. Here,
$\lambda$ denotes the transition wavelength of the molecules. To
decipher details at the resolution of $W$, the detected fluorescence
signal should show a clear change as the antenna is scanned across
an area of the order of $W^2$. The resulting ratio of the signal to
the shot noise of the background becomes $3W^2\Phi\sqrt{
S_0\delta}/\lambda$ if one assumes a numerical aperture of
$N.A.=1.5$. Therefore, if we neglect detector noise, achieving a
signal-to-noise ratio (SNR) of 1 requires $\Phi W\gtrsim \lambda
/(3W\sqrt{S_0 \delta})$. Setting $\delta=1/W^2$, provides a direct
guideline $\Phi W\gtrsim \lambda /3\sqrt{S_0}$ for resolving single
molecules spaced by $W$.

The symbols in Figure~\ref{ex-theory}e display the experimental
results of $\Phi W$ corresponding to the highest and lowest observed
$W$ values reported in Figure~\ref{ex-theory}d, and the solid curves
show the theoretical values for two examples of molecular depth $d$.
We find that the detection SNR is maximized when the GNP has a
diameter of about 80~nm and the attainable resolution is in the
20-30~nm range (see Figure~\ref{ex-theory}d). Although the
experimental data for $\Phi W$ mirrors the same trend, their
accuracy is not sufficient for a quantitative agreement with the
theoretical predictions of the dependence on $d$. To assess the
potential of antenna-based microscopy for ultrahigh resolution
fluorescence imaging, we consider a realistic signal of
$S_0\thickapprox1000-10000$ counts per second from a single
fluorescent molecule at ambient condition and $\lambda=600$~nm,
yielding $\lambda /3\sqrt{S_0}\thickapprox7-2$~nm after an
integration time of one second. Thus, the data in
Figures~\ref{ex-theory}d and \ref{ex-theory}e show that it should be
readily possible to resolve neighboring molecules with spacings
below 20~nm even without the need for any fluorescence background
suppression method~\cite{Xie:06,Hoeppener:09} although such
techniques improve the SNR further.

In conclusion, we have shown that single gold nanoparticles serve as
versatile and well-defined optical nanoantennas for high-resolution
near-field fluorescence microscopy. By varying the particle size, we
investigated the performance of this technique with respect to the
overall attainable fluorescence enhancement and spatial resolution;
we expect our findings to provide also a general trend for other
optical nanoanntennas~\cite{Taminiau:07}. The reported experimental
and theoretical data are in very good agreement and emphasize the
near-field sensitivity of antenna-based microscopy to the exact
position of the fluorophore. Furthermore, rigorous numerical
calculations have shown that the dielectric-air interface imposed by
the sample is responsible for providing particularly high excitation
enhancements and low quenching as compared to a molecule-antenna
assembly embedded in a homogeneous medium. Our analysis shows that
single emitters can be detected at a resolution beyond 20~nm even in
dense ensembles. By using ellipsoidal
nanoparticles~\cite{Mohammadi:08}, it might be possible to push the
resolution in fluorescence microscopy to the molecular scale.

\section*{Acknowledgements}
We thank Mario Agio for fruitful discussion and Alois Renn for
experimental help. This work was supported by ETH Zurich and the
Swiss National Science Foundation (SNF).



\begin{thebibliography}{28}
\bibitem{note2}
\bibinfo{note2}{Treatment of optical spectra at interfaces is a nontrivial task, but many phenomena can be approximated by assuming a homogeneous medium with an effective refractive index.
Here, we set $n=1.35$ as in Ref. \cite{Kuehn:06}, where the
surrounding index was determined by fitting the GNP plasmon spectrum
in that experiment. }
\bibitem{note}
\bibinfo{note}{For dipole moments perfectly perpendicular to the interface, a quantum efficiency of 0.95 is reduced to 0.67. Since our bor-FDTD calculations of the decay rates in Fig. 3b assume such molecules, the results of the fluorescence enhancement are slightly overestimated.}
\end{thebibliography}
\end{document}